\documentclass[twocolumn,showpacs,amsmath,amssymb,prb]{revtex4}
\usepackage{graphicx}
\begin{document}
\draft
\title{
A quantum coherent approach to transport and noise in double  barrier resonant diodes: shot noise a way to distinguish
coherent from sequential tunneling }
\author{V. Ya. Aleshkin}
\address{Institute for Physics of Microstructures, Nizhny Novgorod GSP-105
603600, Russia}
\author{L. Reggiani}
\address{National Nanostructure Laboratory of INFM, Dipartimento di
Ingegneria dell' Innovazione, \\
Universit\`a di Lecce, Via Arnesano s/n, 73100 Lecce, Italy}
\author{N.V. Alkeev, V.E. Lyubchenko}
\address{Institute of Radioengineering
\& Electronics, Russian Academy of Sciences}
\author{C.N. Ironside, J.M.L. Figueiredo and C.R. Stanley}
\address{Department of Electronic and Electrical Engineering,
University of Glasgow}
\date{\today}
%\maketitle
%\Setdate{}
%\TitlePage{}
\vskip 1cm\noindent
\begin{abstract}
{We implement a quantum approach which includes long range Coulomb interaction
and investigate current voltage characteristics and
shot noise in double barrier resonant diodes.
Theory applies to the region of low applied voltages
up to the region of the current peak and considers the wide
temperature range from zero to room temperature. The shape of the
current voltage characteristic is well reproduced and we confirm
that even in the presence of Coulomb interaction shot noise can be
suppressed with a Fano factor well below the value of 0.5. This
feature is a signature of coherent tunnelling since the standard
sequential tunnelling predicts in general a Fano factor equal to
or greater than the value 0.5. This giant suppression  is a
consequence of Pauli principle as well as long range Coulomb
interaction. The theory generalizes previous findings and is
compared with experiments.}
\end{abstract}
\vskip0.5truecm\noindent
\maketitle
\vspace{0.5truecm}
\narrowtext
\section{Introduction}
Since its realization \cite{tsu73},  the double barrier resonant
diode (DBRD) proved to be an electron device of broad physical
interest because of its peculiar non Ohmic current voltage (I-V)
characteristic. Indeed, after a strong superOhmic increase of
current it exhibits a negative differential conductance and
eventually histeresis effects. \cite{goldman87} Even the shot
noise characteristics of DBRDs are of relevant interest in the sense that
suppressed as well as enhanced shot noise with respect to the full
Poissonian value has been observed (see Ref. [\onlinecite{blanter00}]
for a review on the subject). These electrical and noise features
are controlled by the mechanism of carrier tunneling through the
double potential barriers. The microscopic interpretation of these
features is found to admit a coherent or a sequential tunneling
approach \cite{blanter00}. The coherent approach \cite{chang74}
consists in considering the whole device as a single quantum
system characterized by a transmission coefficient describing
carrier transport from one contact to the other. By contrast, the
sequential approach \cite{luryi85} consists in considering
tunneling through the diode as a two step process where carriers
first transit from one contact into the well, then from the well
to the other contact. The intriguing feature of these two
approaches is that from the existing literature it emerges that
both of them are capable to explain the I-V experimental data as
well as most of the shot noise characteristics. Therefore, to our
knowledge there is no way to distinguish between these two
transport regimes and the natural question whether the tunnelling
transport is coherent or sequential remains an unsolved one.
%Since then, a series of theoretical  analysis of this phenomenon
%appeared in the literature based on both the coherent
%\cite{chen91,martin92,hershfield92,jahan95,runge93,sun99}
%and the sequential tunneling models \cite{chen92,davies92,brown92,chen93,%
%hung93,hershfield93,egues94,iannaccone97,blanter99}.
\par
The coherent approach to shot noise in DBRD has received wide
attention since the first experimental evidence by Li et al
\cite{li90} of shot noise suppression with a minimum value of the
Fano factor $\gamma = S_I/(2qI) =0.5$, here $S_I$ is the low frequency
spectral density of current fluctuations and $q$ the
absolute value of the unit charge responsible of current.
Remarkably, most of the coherent approaches developed so far
\cite{chen91,brown92,hershfield92,runge93,iannaccone97,sun99}
predict a maximum suppression
$\gamma=0.5$ even if there is clear experimental evidence of
suppression below this value \cite{brown92,prazadka97,kuznetsov98}
down to values of $\gamma = 0.25$.\cite{prazadka97,kuznetsov98} To
this purpose, some authors obtained theoretical values of the Fano
factor just below the value of 0.5, Ref. [\onlinecite{euges94}] found
0.45 and Ref. [\onlinecite{jahan95}] 0.38, respectively. However, the
physical interpretation of these results remains mostly
qualitative and quoting Ref. [\onlinecite{blanter00}] this direction of
research looks promising but certainly requires more efforts.
We remark that the theory of shot noise in DBRDs
under the sequential approach
\cite{chen92,davies92,chen93,hung93,hershfield93,%
euges94,iannaccone97,blanter99,aleshkin01a,aleshkin01}
provides a Fano factor of $0.5$ as the minimum value of shot noise suppression.
\par
The aim of this paper is to develop a coherent theoretical
approach for current voltage and electronic noise in DBRDs
accounting for Pauli principle and long range Coulomb interaction
going beyond existing models. To this purpose, we implement
the quantum approach proposed in Ref.
[\onlinecite{averin93,blanter99,wei99}].
We anticipate that the main result of the present
theoretical approach concerns with the prediction that suppression
of shot noise  with a Fano factor below 0.5 is a signature of
coherent tunneling against sequential tunneling.
%Such a prediction
%is confirmed by experimental results.
\par
The content of the paper is organized as follows. Section 2
presents the theoretical model. Section 3 provides the analytical
expressions for the calculation of the current voltage
characteristics in the low voltage region limited to the first
peak of the current. Section 4 provides the analytical expression
for the electron noise corresponding to the current voltage
characteristics of Sect. 3. Here, the Nyquist expression is
recovered at vanishing applied voltage. At increasing voltages
suppressed shot noise is found in the region preceding the current
peak and enhanced shot noise at voltages just above the current
peak. Section 5 reports a comparison between theory and existing
experiments. Major conclusions are drawn in Sect. 6.
\section{Model}
The typical diode investigated here is the standard
%symmetric
double well structure depicted in Fig. 1. We denote by $\Gamma =
(\Gamma_{L}+\Gamma _{R})$ the resonant states width, by $\varepsilon_{r}$
the energy of the resonant level as measured from the center
of the potential well and  by $\Gamma _{L,R} $  the partial
widths due to the tunnelling through the left and the right barrier,
respectively. We consider the case of coherent tunnelling in the presence
of only one resonant state and we assume that the diode
has contacts with unit square surfaces.
For convenience, calculations are carried out using the cgs system.
\par
The kinetic model is developed by assuming that the electron
distribution functions in the emitter and in the collector,
$f_\alpha$, are equilibrium-like, but with different
electro-chemical potentials $F_\alpha$:
\begin{equation}
f_{\alpha}(\varepsilon ,F_{\alpha})=\frac{1}{1+\exp \left(
\frac{\varepsilon -F_\alpha} {k_{B}T}\right) }, \label{1}
\end{equation}
here $\alpha=L$ stands for the left contact (the emitter),
$\alpha=R$ for the right contact (the collector), $k_{B}$ is the
Boltzmann constant, $T$ the bath temperature and $\varepsilon $
the  carrier energy.
\par
The full Hamiltonian of the structure is
\begin{equation}
H=H_L+H_R+H_{res}+H_{tun} \label{2}
\end{equation}
here
$$
H_{\alpha}=\sum_{p_\perp}\sum_{p_\alpha}(E(p_\alpha)+p^2_\perp/2m)
c^+_{\alpha}(p_\alpha,p_\perp) \ c_{\alpha}(p_\alpha,p_\perp)
$$
are the Hamiltonians of the right and left contacts, $c^+_{\alpha}$ and $c_{\alpha}$ the creation and annihilation operators of electrons in contact $\alpha$, $p_\perp$ and $p_\alpha$ the momentum components perpendicular and
parallel to the direction of the current, respectively, $m$ is the effective
electron mass in the conduction band, $E(p_\alpha)+p^2_\perp/2m$
the electron energy in the $\alpha$ contact,
$$
H_{res}=\sum_{p_\perp}(E_r+p^2_\perp/2m)a^+(p_\perp) \ a(p_\perp)
$$
is the Hamiltonian of the resonant states, $E_r=\varepsilon_r -qu$
with $u$ the voltage drop between the emitter and center of the
quantum well, $-q$  the electron charge, $a^+$ and $a$ the creation and annihilation operators of electrons in the resonant level, and
$$
H_{tun}=\sum_{p_\perp, \alpha}\left(T_\alpha
a^+(p_\perp)\sum_{p_\alpha}c_\alpha(p_\alpha,p_\perp)+hc\right)
$$
the part of the Hamiltonian describing the electron tunneling,
$T_\alpha$ the amplitude of tunneling between the resonant state
and the $\alpha$th contact.

Following Ref. [\onlinecite{averin93}], the relation between
$\Gamma_\alpha$ and $T_\alpha$ is:
$\Gamma_\alpha =2\pi |T_\alpha|^2 \rho_\alpha$
where $\rho_\alpha$ is the one dimensional density of states, and
the electron operators $a$ and $c$ in the Heisenberg representation are given by:
\begin{equation}
a(t,p_\perp)=\sum_{\alpha, p_\alpha} T_{\alpha}
\frac{\exp(-i[E(p_\alpha)+p^2_\perp/2m]
t/\hbar)}{E(p_\alpha)-E_r+i\Gamma/2}
c_{\alpha}(p_\alpha,p_\perp)\label{3}
\end{equation}
$$
c_{\alpha}(t,p_\alpha,p_\perp)=
c_{\alpha}(p_\alpha,p_\perp)\exp\left(-i[E(p_\alpha)+p^2_\perp/2m]
t/\hbar\right)-
$$
\begin{equation}
-\frac{iT^*_{\alpha}}{\hbar}\int_{-\infty}^{t}d\tau
a(\tau,p_\perp)\exp\left(-i[E(p_\alpha)+p^2_\perp/2m](t-\tau)/\hbar\right)
\label{4}
\end{equation}
%
%where $\Gamma=\Gamma_L+\Gamma_R$.
\par
The current operators for the left and right
contacts $I_{L,R}$ and for the total current $I$ are\cite{averin93}:
\begin{equation}
I_{\alpha}(t)=-\frac{iq}{\hbar}\sum_{p_\alpha,p_\perp}  \left(
T_{\alpha}a^+(t,p_\perp)c_{\alpha}(t,p_{\alpha},p_\perp) -hc
\right)\label{5}
\end{equation}
\begin{equation}
I=\eta I_L-(1-\eta)I_R\label{6}
\end{equation}
where $\eta=u/V$,  $V$ being the total applied voltage (see Fig. 1).

From Eqs. (3) to (5) we obtain the expressions for current
operators similar to that obtained in Ref. [\onlinecite{blanter99}]
$$
I_{\alpha}(t)=\frac{q}{2\pi\hbar}\sum_{p_\perp}\sum_{\beta,
p_\beta}\sum_{\gamma,
p_\gamma}\exp(i(E_\beta-E_\gamma)t/\hbar)\times
$$
\begin{equation}
A^\alpha_{\beta\gamma}(E_\beta,E_\gamma)c^+_\beta(p_\beta,p_\perp)
c_\gamma(p_\gamma,p_\perp)
\end{equation}
where $E_\alpha=E(p_\alpha)$ and the full expression for
$A^\alpha_{\beta \gamma}(E_\beta,E_\gamma)$ is reported in
the Appendix.
From Eqs. (6) and (7) and the expression of
$A^\alpha_{\beta,\gamma}(E_\beta,E_\gamma)$ from the Appendix, we find
the usual expression for the average current
$$
<I>=<I_L>=-<I_R>=
$$
\begin{equation}
=\frac{qm}{2\pi^2\hbar^3}\int_{-g_L}^\infty d\varepsilon_z
D(\varepsilon_z)\int_0^\infty d\varepsilon_\perp
[f_L(\varepsilon,F_L)-f_R(\varepsilon,F_R)]
\end{equation}
where  $\varepsilon _{\bot }=p^2_\perp/2m$ is the kinetic energy of
the transverse motion, $\varepsilon =(\varepsilon _{z}+
\varepsilon_{\bot })$, $g_{L}\left( u_{L}\right) $ the energy gap
between the bottom of the conduction band and the first
quantized level in the well before the left barrier (see Fig.
1), $D(\varepsilon_z)$ the transparency of the double barrier given by:
$$
D(\varepsilon_{z}) = \frac{\Gamma_L \Gamma_R }{(\varepsilon_{z}
-\varepsilon _{r}+qu)^{2}+\frac{\Gamma ^{2}}{4}}
$$
In our model, we suppose that in the emitter there are no electron
states with energy below $-g_L$ and that electron states with
energy higher than this value are three dimensional.
\par
To take into account Coulomb effects, we introduce the expression for
the operator of the electron charge in the quantum well, that from Eq. (3)
is found to be given by:
$$
Q_{QW}(t)=-q\sum_{p_\perp}a^+(t,p_\perp)a(t,p_\perp)=
$$
$$
=-q\sum_{p_\perp}\sum_{\alpha,p_\alpha}\sum_{\beta,
p_\beta}\frac{T^*_\alpha T_\beta \exp(i(E_\alpha-E_\beta)t/\hbar)}
{(E_\alpha-E_r-i\Gamma/2)(E_\beta-E_r+i\Gamma /2)}\times
$$
\begin{equation}
\times c^+_\alpha(p_\alpha,p_\perp)c_\beta(p_\beta,p_\perp)
\end{equation}\label{9}
From Eq. (9) the average electron charge in the quantum well is found
to be:
$$
<Q_{QW}>=-\frac{qm}{2\pi^2\hbar^2}\left[ \int_{-g_L}^\infty
d\varepsilon_z D(\varepsilon_z)\int_0^\infty d\varepsilon_\perp
\Gamma_R^{-1}f_L(\varepsilon, F_L) \right.
$$
\begin{equation}
+\left.\int_{-qV}^\infty d\varepsilon_z
D(\varepsilon_z)\int_0^\infty d\varepsilon_\perp
\Gamma_L^{-1}f_R(\varepsilon, F_R) \right].
\end{equation}\label{10}
\section{Current voltage characteristics}
The current voltage characteristic is determined from Eq. (8)
once the dependence of $D(\varepsilon_z)$ and $g_L$ on the applied voltage
is given.
To calculate the transparency explicitly, we must find $u_{L}(V)$
and $u(V)$ as functions of $V$. To this
purpose, we consider the Poisson equation for the electrical
potential $\varphi$ in the structure  of Fig. 1.
In the emitter, the Poisson equation can be written in the form
\begin{equation}
\varphi''=\frac{4\pi q}{\kappa
%\kappa_0
 }\left[ N_{c}F_{1/2}\left( \frac{%
F_{L}+q\varphi }{k_{B}T}\right) -n\right] \label{11}\end{equation}
with
$$
n=N_{c}F_{1/2}\left( \frac{F_{L}}{k_{B}T}\right)
$$
the electron concentration in the emitter,
$N_{c}$ the effective density of states of the conduction band,
$F_{1/2}(x)$ the Fermi-Dirac integral of index 1/2~
\cite{ashcroft76}, $\kappa$ the static dielectric constant
of the material.
We note that the effect of size quantization on the electron
distribution in the emitter is neglected. By integrating Eq.
(11) and taking into account that $\varphi \left( -\infty \right)
=\varphi'\left( -\infty \right) =0$, we find the relation between
$ \varphi _{L}'$ and  $u_{L}$  on the left border of the left
barrier
$$
\varphi '_{L}=\sqrt{\frac{8\pi k_B T}{\kappa
%\kappa_0
}\times}
$$
\begin{equation}
\left[ N_{c}F_{3/2}\left( \frac{ F_{L}+qu_{L}}{k_{B}T}\right)
-N_{c}F_{3/2}\left( \frac{F_{L}}{k_{B}T}\right) -\frac{q}{k_B
T}nu_{L}\right]^{1/2 } \label{12}
\end{equation}
Here $F_{3/2}$ is the Fermi-Dirac integral of index
3/2.\cite{ashcroft76} To simplify the task, we suppose that the
barriers are undoped and that the charge in the quantum well is
placed at its middle plane, so that we can write
\begin{equation}
u=u_{L}+\varphi '_{L}\left( d_{L}+d_{QW}/2\right) \label{13}
\end{equation}
and
\begin{equation}
u_{R}=u+\left( \varphi '_{L}-\frac{4\pi }{\kappa
%\kappa_0
}Q\right) \left( d_{R}+d_{QW}/2\right) \label{14}
\end{equation}
with $d_{L}$, $d_{R}$, $d_{QW}$ the widths of, respectively,
the left battier, the right barrier, and the quantum well;
$Q=(qN_{QW}^{+}-Q_{QW})$ the charge in the quantum well;
$N_{QW}^{+}$  the number of charged donors  in the quantum well.
Furthermore, we  suppose that the electron concentration in the
collector is the same of that in the emitter, hence $F_{R}=(F_{L}-qV)$ and the Poisson equation in the collector takes
the  form given by Eq. (\ref{11}) with the change
$F_{L}\rightarrow F_{R}$.
By integrating this new equation with
the condition $\varphi \left( \infty \right) =V$, $\varphi '\left(
\infty \right) =0$, in analogy with Eq. (12) we obtain
$$
-\frac{k}{4\pi }\varphi'_{L}+Q+\frac{k}{4\pi }\sqrt{\frac{8\pi k_B
T}{ \kappa }}
 \left[ N_{c}F_{3/2}\left(
\frac{F_{L}+qu_{R}-qV}{k_{B}T}\right)\right.
$$
\begin{equation}
\left.-N_{c}F_{3/2}\left( \frac{F_{L}-qV}{k_{B}T}\right)
-\frac{qn}{k_B T}\left( u_{R}-V\right) \right]^{1/2}=0 \label{15}
\end{equation}
Equation (\ref{15}) relates  $u_{R}$ to $V$. We remark that it is
more convenient to consider $V$ and $<I>$ as functions of $u$
because in this case they are single valued functions even for I-V
characteristics of Z-type. Since the first and the third term in
the left hand side of Eq. (\ref{15}) are the charge of the emitter
$Q_{L}$ and of the collector $Q_{R}$, respectively, Eq. (\ref{15})
expresses the electroneutrality  condition of the device. We
note, that to derive Eq. (\ref{15})  we have assumed that the
width of the depletion region in the collector region (see Fig. 1)
is smaller than that between the right barrier and the highly
doped region in the collector. As typical in  DBRDs, from both
sides of the barriers there are spacers with doping level of the
order of $10^{16} \div 10^{17} \ cm^{-3}$ and with widths in the range
between $10 \div 500 \ nm$ \cite{brown92,figueiredo99}.
If the low doped region in the collector is fully
depleted, then Eq. (\ref{13}) must be substituted  with
\begin{equation}
V=u_R+\varphi'_R L-\frac{2\pi q n L^2}{\kappa
%\kappa_0
} \label{16}
\end{equation}
where $L$ is the width of the spacer in the collector, and
$\varphi'_R=(\varphi'_L-4\pi Q_{QW}/\kappa$).
In the  derivation of Eq. (\ref{16}), the voltage drop in the
highly doped collector region is neglected.
\section{Noise}
To calculate the spectral density of current fluctuations at zero
frequency under fixed voltage we use the expression\cite{averin93}:
\begin{equation}
S_I(0)=\int_{-\infty}^\infty dt <\delta I(0)\delta I(t)+\delta
I(t)\delta I(0)>
\end{equation}
We anticipate that the total current fluctuation consists of
two sources.
The first comes from the fluctuation of population states in the
contacts and the second from the fluctuation of the electron charge in
the quantum well \cite{blanter99}.
This last leads to fluctuations of the voltage
drop between the emitter and the quantum well, $\delta u$.
Accordingly, the operator of current fluctuations is given by:\cite{blanter99}
\begin{equation}
\delta I(t)=\delta I_2(t)+\frac{\partial <I>}{\partial u}\delta
u(t)
\end{equation}
where $\delta I_2(t)$ is the current fluctuation operator
under fixed $u$ due to the population fluctuations in the contacts.
Analogously we can introduce the operator of charge fluctuation
\begin{equation}
\delta Q_{QW}(t)=\delta Q_{QW2}(t)-C_{QW}\delta u(t)
\end{equation}
where $C_{QW}$ is the differential capacitance of the quantum well
$$
C_{QW}=-\frac{\partial <Q_{QW}>}{\partial u}.
$$
From the condition of charge neutrality,
the charge fluctuations in the emitter $\delta Q_e$, in the
collector $\delta Q_c$, and in the quantum well $\delta Q_{QW}$ satisfy:
\begin{equation}
\delta Q_e+\delta Q_c+\delta Q_{QW}=0
\end{equation}
The charge fluctuation in the emitter and collector can be expressed
through $\delta u_{L,R}$ as:
\begin{equation}
\delta Q_e=-C_e\delta u_L,\quad \delta Q_c=-C_c\delta u_R
\end{equation}
where $C_{e,c}=-\partial Q_{e,c}/\partial u_{L,R}$ ($V=const$) are,
respectively, the differential capacitance of the charge in the
accumulation region of the emitter and in the depletion region
of the collector whose expressions are detailed in the Appendix
(we note that when Eq. (16) is valid $C_c=\kappa/(4\pi L)$).
By taking in account that
$\varphi'_L=-4\pi Q_e/\kappa$, from Eq. (13) we find:
\begin{equation}
\delta u= \delta u_L (1+C_e/C_1),\quad
C_1=\frac{\kappa}{4\pi(d_L+d_{QW}/2)}
\end{equation}
Analogously, by taking into account that $\varphi'_L-4\pi
Q/\kappa=4\pi Q_c/\kappa$, from Eq. (14) the relation between
$\delta u$ and $\delta u_R$ is found to be:
\begin{equation}
\delta u= \delta u_R(1+C_c/C_2),\quad
C_2=\frac{\kappa}{4\pi(d_R+d_{QW}/2)}
\end{equation}
From Eqs. (21) to (23), Eq. (20) is rewritten in the form
\begin{equation}
\delta Q_{QW}=\delta u (C_L+C_R),\quad
C_{L,R}=\frac{C_{e,c}C_{1,2}}{(C_{e,c}+C_{1,2})}
\end{equation}
where $C_L$, ($C_R$) is the capacitance  between the emitter
(collector) and the center of the quantum well, respectively.
From Eq. (24) it is clear that $C_{L,R}$ can be considered as two
capacitances in series $C_{1,2}$ and $C_{e,c}$. If we neglect
the accumulation and depletion regions (i.e. $C_{e,c}\rightarrow \infty$)
and neglect the quantum well width (i.e. $d_{QW}\rightarrow 0$) then
$C_{L,R}$ coincide with those used in Ref. [\onlinecite{blanter99}].
From Eqs. (19) and (24), the relation between $\delta u$ and $\delta
Q_{QW2}$ is found to be:
\begin{equation}
\delta u(t) =\frac{\delta Q_{QW2}(t)}{C_L+C_R+C_{QW}}
\end{equation}
From Eqs. (17), (18) and (25), $S_I(0)$ is found to be given by the sum
of three terms as:
$$
S_I(0) = S_1 + S_2 + S_3
$$
The first, $S_1$, is the correlator of $\delta I_2$;
the second, $S_2$, is proportional to the cross correlator between
$\delta I_2$ and $\delta Q_{QW2}$;
the third, $S_3$, is proportional to the correlator of $\delta Q_{QW2}$.
They are given explicitely by:
\begin{equation}
S_1=\int_{-\infty}^\infty dt<\delta I_2(0)\delta I_2(t)+ \delta
I_2(t)\delta I_2(0)>
\end{equation}
$$S_2=J\int_{-\infty}^\infty dt<\delta I_2(0)\delta Q_{QW2}(t)+
\delta I_2(t)\delta Q_{QW2}(0)+
$$
\begin{equation}
+\delta Q_{QW2}(0)\delta I_2(t)+\delta Q_{QW2}(t)\delta I_2(0)>
\end{equation}
$$
S_3=J^2\int_{-\infty}^\infty dt<\delta Q_{QW2}(0)\delta
Q_{QW2}(t)+$$
\begin{equation} + \delta Q_{QW2}(t)\delta
Q_{QW2}(0)>
\end{equation}
where
$$
J=\frac{1}{C_L+C_R+C_{QW}}\frac{\partial <I>}{\partial u}
$$
plays the role of a differential RC rate.
\par
By using the definitions for $\delta I_{\alpha 2}(t)$
$$
\delta I_{\alpha 2}(t)=
\frac{q}{2\pi\hbar}\sum_{p_\perp}\sum_{\beta,
p_\beta}\sum_{\gamma,
p_\gamma}\exp(i(E_\beta-E_\gamma)t/\hbar)A^\alpha_{\beta\gamma}
(E_\beta,E_\gamma)\times
$$
\begin{equation}
\left(c^+_\beta(p_\beta,p_\perp)c_\gamma(p_\gamma,p_\perp)-
<c^+_\beta(p_\beta,p_\perp)c_\gamma(p_\gamma,p_\perp)>\right)
\end{equation}
for $\delta Q_{QW2}(t)$
$$
\delta
Q_{QW2}(t)=-q\sum_{p_\perp}\sum_{\alpha,p_\alpha}\sum_{\beta,
p_\beta}\left(
c^+_\alpha(p_\alpha,p_\perp)c_\beta(p_\beta,p_\perp)\right.-
$$
$$
-\left.<c^+_\alpha(p_\alpha,p_\perp)c_\beta(p_\beta,p_\perp)>\right)\times
$$
\begin{equation}\times \frac{T^*_\alpha T_\beta \exp(i(E_\alpha-E_\beta)t/\hbar)}
{(E_\alpha-E_r-i\Gamma/2)(E_\beta-E_r+i\Gamma /2)}\end{equation}
and the property \cite{blanter99}
$$
<c^+_\alpha(p_\alpha,p_\perp)c_\beta(p_\beta,p_\perp)c^+_\gamma(p_\gamma,p'_\perp)
c_\delta(p_\delta,p'_\perp)>-
$$
$$
-<c^+_\alpha(p_\alpha,p_\perp)c_\beta(p_\beta,p_\perp)><c^+_\gamma(p_\gamma,p'_\perp)
c_\delta(p_\delta,p'_\perp)>=
$$
$$
=\delta_{\alpha \delta}\delta_{p_\alpha p_\delta}\delta_{\beta
\gamma}\delta_{p_\beta p_\gamma}\delta_{p_\perp
p'_\perp}f_\alpha(E_\alpha+p^2_\perp/2m, F_\alpha)\times
$$
\begin{equation}
\times\left(1-f_\beta(E_\beta+p^2_\perp/2m, F_\beta)\right)
\end{equation}
it is possible to find  for $S_i$ $i =1,2,3$ the expressions
$$
S_1=\frac{q^2m}{\pi^2\hbar^3}\int_{-g_L}^{\infty}d\varepsilon_z\int_0^\infty
d\varepsilon_\perp\left\{D \left[f_L(1-f_R)+\right.\right.
$$
\begin{equation}
\left.\left.+f_R(1-f_L)\right]-D^2 (f_L-f_R)^2\right\}\label{32}
\end{equation}
$$
S_2=-\lambda\frac{q^2m}{\pi^2\hbar^3}\int_{-g_L}^{\infty}d\varepsilon_z\int_0^\infty
d\varepsilon_\perp
D^2\left\{\frac{2\Gamma_L}{\Gamma}f_L(1-f_L)\right.
$$
\begin{equation}
\left.-\frac{2\Gamma_R}{\Gamma}f_R(1-f_R)+\frac{(\Gamma_R-\Gamma_L)}{\Gamma}
[f_L(1-f_R)+f_R(1-f_L)]\right\}\label{33}
\end{equation}
$$
S_3=\lambda^2\frac{q^2m}{\pi^2\hbar^3}\int_{-g_L}^{\infty}d\varepsilon_z\int_0^\infty
d\varepsilon_\perp D^2 \times
$$
$$
\times\left\{
\frac{\Gamma_L^2}{\Gamma^2}f_L(1-f_L)+\frac{\Gamma_L\Gamma_R}{\Gamma^2}
[f_L(1-f_R)+f_R(1-f_L)]\right\}+
$$
\begin{equation}
+\lambda^2\frac{q^2m}{\pi^2\hbar^3}\int_{-qV}^{\infty}d\varepsilon_z\int_0^\infty
d\varepsilon_\perp
D^2 \frac{\Gamma_R^2}{\Gamma^2}f_R(1-f_R)\label{34}
\end{equation}
Here we used the notation $f_{L,R}=f_{L,R}(\varepsilon, F_{L,R})$
and
$$
\lambda=\frac{\hbar \Gamma}{\Gamma_L\Gamma_R}\frac{1}{(C_L+C_R+C_{QW})}
\frac{\partial <I>}{\partial u}.
$$
where $- \infty < \lambda < \infty$ is a dimensionless parameter
describing Coulomb interaction
\par
Equations (32)-(34) are the central result of this paper. We note
that $S_I(0)$  does not depend on $\eta$. When the $\lambda=0$,
i.e. Coulomb interaction is negligible, $S_I(0)=S_1$ and the
result of Ref. [\onlinecite{martin92}] is recovered.
\par
As an internal check we prove that $S_I(0)$ satisfies
Nyquist theorem\cite{nyquist25}.
Indeed, at zero applied voltage $f_{L}=f_{R}$,
from the expression (8) for the total current
it follows that $\partial <I>/\partial u=0$, and therefore,
$\lambda=0$.
Accordingly, the differential conductance $G$ at zero voltage is
\begin{equation}
G=\frac{q^2m}{2\pi \hbar^3 k_B T}\int_0^\infty d\varepsilon_z
D\int_0^\infty d\varepsilon_\perp f_L(1-f_L)
\end{equation}
We here used the property that for $V=0$
$\partial f_R/\partial qV=f_L(1-f_L)/(k_BT)$.
According to Eq. (32), $S_I$ at zero voltage is
\begin{equation}
S_I(0)=\frac{2q^2m}{\pi \hbar^3}\int_0^\infty d\varepsilon_z
D\int_0^\infty d\varepsilon_\perp f_L(1-f_L)
\end{equation}
Equations (35) and (36) imply  $S_I(0)=4k_BTG$, which represents
Nyquist theorem.\par
Let us now show that $S_I(0) \rightarrow \infty $ on the border of
the instability region where $ d<I>/dV \rightarrow \infty $.
To this purpose, we note that $dI/dV$ can be decomposed as:
\begin{equation}
\frac{d<I>}{dV}=\frac{\partial <I>}{\partial V}+\frac{\partial
<I>} {\partial u} \frac{du}{dV} \label{37}
\end{equation}
By writing the condition of charge neutrality as:
\begin{equation}
-C_edu_L+dQ_{QW}+C_c(dV-du_R)=0
\end{equation}
by taking into account Eqs. (22), (23) and that
\begin{equation}
dQ_{QW}=-C_{QW}du+\frac{\partial N_{QW}}{\partial V}dV
\end{equation}
we obtain
\begin{equation}
\frac{du}{dV}=\left(C_c+\frac{\partial Q_{QW}}{\partial
V}\right)(C_L+C_R+C_{QW})^{-1}
\end{equation}
By using Eq. (8) for $<I>$, we see that $\partial <I>/\partial
u_L$ and $\partial <I>/\partial V$ entering Eq. (37) are always
finite. This  implies that $d<I>/dV  \rightarrow \infty $  only if
the sum $(C_L+C_R+C_{QW}) \rightarrow 0$. We remark that also the
denominator of $\lambda$ is proportional to $(C_L+C_R+C_{QW})$;
thus, $\lambda$, and in turn $S_I(0)$ go to infinity
simultaneously with $dI/dV$. Note that $C_{L,R}$ are always
positive and $C_{QW}$ becomes negative when the resonant state
approaches $g_L$, that is  when the number of electrons in the
quantum well decreases at increasing  $u$.
\section{Results and discussion}
In this section  we present the theoretical results
for the two relevant cases of zero and high temperatures
(i.e. $T \gg \Gamma$ which here corresponds to $T \ge 77 \ K$),
the former being appropriate to investigate the influence on shot noise
of the Pauli principle and the latter of long
range Coulomb interaction.
Then, theoretical results are compared with experiments.
\subsection{Zero temperature. Three dimensional case}
We investigate the condition of high applied voltages, when
$qV > F_R$, because more close to experiments.
(We recall that typical  magnitudes for the relevant parameters of
DBRDs are: for $\Gamma $ less than a few $meV$, for $F_L$ less than
$100 \ meV$,
and for the voltage at the  peak current around $0.5 \ V$.)
Because of the above, from Eq. (8) the expression for the current becomes:
\begin{equation}
<I>=\frac{qm}{2\pi^2\hbar^3}\int_{-g_L}^{F_L}d\varepsilon
(F_L-\varepsilon)D(\varepsilon)=\frac{qm\Gamma_L\Gamma_R}{4\pi^2\hbar^2}B(f,\xi)
\end{equation}
where
$$
B(f,\xi)=2(f+\xi)\left[\arctan(f+\xi)-\arctan(\xi)\right]-$$
$$-\ln\left(\frac{1+(f+\xi)^2}{1+\xi^2}\right),$$
and for convenience we define dimensionless chemical potential $f$ and voltage
drop $\xi$ as
$$
f=\frac{2(F_L+g_L)}{\Gamma},\quad
\xi=\frac{2(qu-\varepsilon_r-g_L)}{\Gamma}
$$
The expression for the noise spectral density becomes
$$
S_I(0)=\frac{q^2m}{\pi^2\hbar^3}\int_{-g_L}^{F_L}d\varepsilon
(F_L-\varepsilon)D(\varepsilon)[1-D(\varepsilon)]+
$$
\begin{equation}
+\frac{q^2m}{\pi^2\hbar^3}\left[\lambda\frac{(\Gamma_L-\Gamma_R)}{\Gamma}
+\lambda^2\frac{\Gamma_L\Gamma_R}{\Gamma^2}\right]\int_{-g_L}^{F_L}d\varepsilon
(F_L-\varepsilon)D(\varepsilon)^2
\end{equation}
For the Fano factor, $\gamma=S_I(0)/2q<I>$, Eqs. (41) and (42) yield:
\begin{equation}
\gamma=1-\frac{4\Gamma_L\Gamma_R}{\Gamma^2}\left\{1-\lambda
\frac{(\Gamma_L-\Gamma_R)}{\Gamma}-\lambda^2\frac{\Gamma_L\Gamma_R}{\Gamma^2}\right\}
\frac{A(f,\xi)}{B(f,\xi)}
\end{equation}
where
$$
A=(f+\xi)\left[\frac{f+\xi}{1+(f+\xi)^2}+\arctan(f+\xi)-
\frac{\xi}{1+\xi^2}\right.
$$
$$
\left.-\arctan(\xi)\right]\\
 +\frac{1}{1+(f+\xi)^2}-\frac{1}{1+\xi^2},
$$
%From Eq. (43) we found convenient to analyze the
%dependence of $\gamma$ on $u$ (more exactly on $\xi$) instead of the
%total voltage $V$.
We note that $\lambda$ depends on $f$ and $\xi$ through $C_{L,R}$,
$C_{QW}$, $\partial <I>/\partial u$ which are given by Eq. (24) and
by:
\begin{equation}
C_{QW}=\frac{q^2m}{\pi^2\hbar^2}\frac{\Gamma_L}{\Gamma}H(f,\xi),\quad
\frac{\partial<I>}{\partial
u}=\frac{q^2m\Gamma_L\Gamma_R}{\pi^2\hbar^3\Gamma}H(f,\xi)
\end{equation}
$$
H(f,\xi)=\left[\arctan(f+\xi)
-\arctan(\xi)-\frac{f}{\xi^2+1}\right]+
$$
$$
+\frac{\Gamma^2}{4\Gamma_L\Gamma_R}\frac{\partial g_L}{\partial
qu}fD(-g_L)
$$
If $f\gg 1$ and the resonant level is located below the the Fermi level
of the emitter far from its borders $F_L$ and $-g_L$, i.e. $\xi\ll
-1$, $f+\xi\gg 1$,  then $A(f,\xi)/B(f,\xi)\approx 1/2$ and Eq. (43)
recovers the expression given in Ref. [\onlinecite{blanter99}]
\begin{equation}
\gamma=\frac{\Gamma_L^2+\Gamma_R^2+2\Lambda(\Gamma_R-\Gamma_L)+2\Lambda^2}{\Gamma^2}
\end{equation}
with
$$
\Lambda=-{\lambda\Gamma_L\Gamma_R \over \Gamma}
$$
As it will be shown below, when $f\gg 1$ the values taken by $\gamma$
in  Eq. (45) are practically constant and correspond to
the plateau exhibited by the dependence $\gamma(f,\xi)$.
On this plateau, $\gamma\ge 1/2$ (the minimum is reached when
$\Lambda=(\Gamma_L-\Gamma_R)/2$). Note that on the plateau
$$
C_{QW}\approx \frac{q^2m\Gamma_L}{\pi\hbar^2\Gamma},\quad
\lambda\approx
\frac{q^2m}{\pi\hbar^2}\frac{1}{C_L+C_R+C_{QW}}<\frac{\Gamma}{\Gamma_L}
$$
and the expression in the braces of Eq. (43) is positive.
Therefore, on the plateau $1\ge \gamma\ge 1/2$ and shot noise
enhancement is impossible.
\par
Now we demonstrate that at voltage values for which the resonant
level is close to the  band edge of the emitter the Fano factor
can drop below the value $1/2$. To this purpose, let us firstly
consider the simplified model where we take $\kappa=12.9$ (GaAs),
$C_L=C_R=\kappa/4\pi d$, $d=5 \ nm$ and neglect the term
proportional to $\partial g_L/\partial u$ in the expression for
$H(f,\xi)$. Figure 2 reports the dependencies of current and Fano
factor on $\xi$ for the two values $f=100$ (Fig. 2 (a)), $f=10$
(Fig. 2(b)) when $\Gamma_L=\Gamma_R$ and in the presence
(continuous curves) or absence (dotted curves) of Coulomb
interaction. The figure shows that for large value of the ratio
between the Fermi energy and the resonant width ($f=100$),
$\gamma(\xi)$ exhibits a wide plateau region  where $\gamma\approx
0.55$ followed by a minimum with $\gamma_{min}\approx 0.464$. By
further increasing the value of $f$ ($f=1000$) the plateau region
is found to widen and  $\gamma_{min}\approx 0.49$.
%The presence of Coulomb interaction is always found to
%enhance the Fano factor.
Note that, as it follows from Eq. (42), the Coulomb interaction
always increases the noise if $\Gamma_L=\Gamma_R$
%and $\lambda=0$ when $\gamma=\gamma_{min}$.
With the decrease of $f$ (see Fig. 2(b) where $f=10$),
the plateau region becomes narrower and  $\gamma_{min}$ is found
to decrease.
The decrease of the value of $\gamma_{min}$ is due to two
complementary reasons.
The first is associated with the decreasing of the strength of Coulomb
interaction.
The second is associated with the increase of the effective barrier
transparency near to the current peak and in turn with the
further suppression of partition noise.
For the ideal case $f\ll 1$, at the peak current the transparency
$D \rightarrow 1$ and in turn $\gamma_{min}\rightarrow 0$.
\par
Why does Coulomb interaction increase the shot noise if
$\Gamma_L=\Gamma_R$ ? To answer this question, we analyze the
noise contribution due to electrons with energies implying $D
\rightarrow 1$. The part of $S_1$ corresponding to these electrons
is zero, while the part of $S_3$ is finite. We remind that the
term $S_1$ describes partition noise, thus it is proportional to
$D(1-D)$ and for $D=1$ at zero frequency there is no noise
associated with the fluctuation of $\delta I_2$. However, even for
$D=1$ there is charge fluctuation in the quantum well  due to the
random character of electron escape from the well (random
time-delay of charge). The probability of this escape is
proportional to $\Gamma$ and this is the reason why $S_3$
decreases with the increase of $\Gamma$. Therefore, Coulomb
interaction enhances current noise due to the random time-delay of
the charge in the quantum well, and which represents the quantum
property of the electron motion in the RTD.
\par
The effects due to the asymmetry of the diode barriers  when
$\Gamma_L=0.25\Gamma$ and $\Gamma_L=0.75\Gamma$ are shown in
Fig. 3 which reports the dependencies of current and Fano factor on $\xi$
for the two values $f=100$, $f=10$.
We note, that without Coulomb interaction (i.e. $\lambda=0$) these
dependencies
%of the current and Fano factor  on $\xi$
are  the same
for $\Gamma_L=0.25\Gamma$ and $\Gamma_L=0.75\Gamma$.
From this figure it is clear that on the plateau Coulomb
interaction decreases the noise  for $\Gamma_L=0.25\Gamma$
while increases the noise for $\Gamma_L=0.75\Gamma$.
This asymmetry is due to the cross correlation term $S_2$ which on the
plateau is positive for $\Gamma_L>\Gamma_R$ and negative for
$\Gamma_L<\Gamma_R$. The physical reason of this asymmetry stems
from the following fact.
If $\Gamma_L>\Gamma_R$, then the charge situated in the quantum well
is characterized by an escaping probability to the emitter
which is greater than that to the collector.
By contrast, when $\Gamma_L<\Gamma_R$ the opposite happens.
Of course, when $\Gamma_L=\Gamma_R$
the escaping probability to the emitter and to the collector is
the same and $S_2=0$. We note the important role played by  Pauli principle for the
possibility of $S_2$ to be positive.
From Eq. (33) it is clear that if
$f_L\ll 1$ and $f_R=0$ then $S_2<0$ for any value of the ratio
$\Gamma_L/\Gamma_R$.
From Fig. 3 one can also see that $\gamma_{min}=0.457<0.5$ for
$\Gamma_L=0.75\Gamma$ and $f=100$ (see Fig. 3(a))
and for both values of $\Gamma_L$ when $f=10$ (see Fig. 3(b)).
\par
We conclude, that for the simple model considered here both
the increase of $\Gamma$ and the decrease of $F_L$ decreases
$\gamma_{min}$ which value drops below 0.5 for $\Gamma_L>\Gamma_R$.
We further note, that there is bias when  $\partial <I>/\partial u=0$ and
$\lambda$ changes of sign. Under this bias, for $\Gamma_L=\Gamma_R$ the
curves of the Fano factor with and without Coulomb interaction touch
each other (see Fig. 2) while for$\Gamma_L\ne\Gamma_R$ they cross
(see Fig. 3).
\par
To confirm the possibility of evidencing the giant suppression
$\gamma_{min}<0.5$ in a real structure,  Fig. 4 presents the
calculations performed with a set of parameters related to the
experiments in Ref. [\onlinecite{brown92}] at $T=4.2 \ K$ and
where both a symmetric (continuous curve) and an asymmetric
(dashed curve) are considered. The structure parameters are $n=2
\times 10^{16} \ cm^{-3}$, $d_L=d_R=d_{QW}=5 \ nm$, squared area
of contacts  $50 \ \mu m^2$ and $L=50 \ nm$, $m=0.067 \ m_0$, with
$m_0$ the free electron mass and $\kappa  =12.9$. In the symmetric
structure, the only two fitting parameters are $\Gamma_L=0.5
\Gamma=0.48 \ meV$ and $\varepsilon_r=104 \ meV$. Their values
control the location and the amplitude of the current peak,
respectively, and are chosen by optimizing the agreement between
the experimental and calculated I-V characteristics at $77 \ K$.
For the asymmetric structure there are three parameters and we
take
%COMMENT SYMMETRIC AND ASYMMETRIC ??
$\Gamma_L=0.25 \Gamma$, $\Gamma=1.22$~meV, $\varepsilon_r=112$~meV
to preserve the location of the current peak at the same voltage.
The function $g_L(u_L)$ is calculated by solving the Schr\"odinger
equation for a potential which: (i)
for $x<0$  follows from the Poisson equation without accounting for
quantization effects; (ii) for $x>0$  corresponds to the solid solution
$Al_{0.42}Ga_{0.58}As$ with zero electric field
inside, as was done in Ref. [\onlinecite{brown92}].
For these values, the dependence $g_L(u_L)$
is found to be almost linear and well approximated by:
$g_L(u_L)\approx 0.28(qu_L-110k_B T)-14 k_B T$.
The details of the solution of the Schr\"odinger equation are
reported  in the Appendix. Calculations give  $\gamma_{min}=0.43$
in reasonable agreement with the value of 0.35 found in experiments \cite{brown92}.
From Fig. 4 we see that both the I-V and the noise characteristics are sensitive
to the asymmetry of the structure, as expected. In any case,
the main features of both characteristics are preserved.
\par
We conclude, that the main reason for shot noise suppression in
RTDs at  temperatures below about $4.2 \ K$ is essentially related
to Pauli principle and, because of the coherent tunneling regime,
near to the current peak the Fano factor can take values
significantly lower than $0.5$ (giant shot-noise suppression).
\subsection{Zero temperature. One dimensional case}
Here we shall discuss shot noise suppression for the case
of one dimensional geometry.
We recall that the possibility for $\gamma_{min}<0.5$ when the
resonance level touches the Fermi level  in the absence of Coulomb
interaction follows from the expression of the noise spectral density
originally derived in Ref. [\onlinecite{averin93}] and further
discussed in Ref. [\onlinecite{wei99}].
At zero temperature  we can write the expressions for the current and
current noise spectral density in the following forms (spin
degeneracy included)
\begin{equation}
<I>=\frac{q}{\pi\hbar}\int_{-g_L}^{F_L}d\varepsilon D(\varepsilon)
\end{equation}
$$
S_I(0)=\frac{2q^2}{\pi\hbar}\left[\int_{-g_L}^{F_L}d\varepsilon
D(\varepsilon)\right.-
$$
\begin{equation}
-\left.\left\{1-\lambda\frac{\Gamma_L-\Gamma_R}{\Gamma}-
\lambda^2\frac{\Gamma_L\Gamma_R}{\Gamma^2}
\right\}\int_{-g_L}^{F_L}d\varepsilon D(\varepsilon)^2\right]
\end{equation}
By using Eqs.  (46) and (47) and after integration,  for $\gamma$we obtain the expression:
\begin{equation}
\gamma(f,\xi)=1-\frac{2\Gamma_L\Gamma_R}{\Gamma^2}\left\{1-\lambda
\frac{\Gamma_L-\Gamma_R}{\Gamma}
-\lambda^2\frac{\Gamma_L\Gamma_R}{\Gamma^2}\right\}\frac{A_1(f,\xi)}{B_1(f,\xi)}
\end{equation}
here
$$
A_1(f,\xi)=\left[\frac{f+\xi}{(f+\xi)^2+1}-\frac{\xi}{\xi^2+1}+
\right.$$$$+\left.\arctan(f+\xi)-\arctan(\xi) \right]
$$
$$
B_1(f,\xi)=\arctan(f+\xi)-\arctan(\xi)
$$
The dependencies $C_{QW}$ and $\partial <I>/\partial u$ on $f,\xi$
are
$$
C_{QW}=\frac{4q^2\Gamma_L}{\pi\Gamma^2}H_1(f,\xi),\quad
\frac{\partial <I>}{\partial
u}=\frac{4q^2\Gamma_L\Gamma_R}{\pi\Gamma^2\hbar}H_1(f,\xi)
$$
with
$$
H_1(f,\xi)=\frac{1}{(f+\xi)^2+1}-\frac{1}{\xi^2+1}+
\frac{\Gamma^2}{4\Gamma_L\Gamma_R}D(-g_L)\frac{\partial
g_L}{\partial qu}
$$
If we neglect the Coulomb interaction (i.e. $\lambda=0$),
introduce $z=f+\xi$ and take $\xi\rightarrow -\infty$ (the so called
infinite band model\cite{jauho94}),  then Eq. (47) recovers
the expression for the noise spectral density derived
in Ref. [\onlinecite{averin93}].
We note, that similarly to the three dimensional case there is a
plateau region in the behavior $\gamma(\xi)$ when $f\gg 1$.
However, in the present one dimensional case $\partial<I>/\partial u$ is
small and hence Coulomb interaction weakly influences the
current noise.
In contrast to the three dimensional case, here the Coulomb
interaction influences the noise  when the resonant level is: either close
to the Fermi level in the emitter $F_L$, or to the conduction band
edge $-g_L$.
We note, that here the expressions for the capacitances derived in the
three dimensional case are no longer valid, and furthermore
we have assumed $C_{L,R}$ to be constant.
Since  when $f\gg 1$ the minimum value of $C_{QW}$ is approximatively
given by $-4q^2\Gamma_L/\pi\Gamma^2$,  for the
appearance of a current instability due to Coulomb interaction
it should be $(C_L+C_R)>4q^2\Gamma_L/\pi\Gamma^2$.
\par
Figure 5 reports the calculated dependencies of the current and
Fano factor versus $\xi$  for the symmetrical structure
$\Gamma_L=\Gamma_R$ when $f=100$ (Fig. 5(a)) and $f=10$ (Fig.
5(b)), respectively. Here, we take $\pi\Gamma (C_L-C_R)/2q^2=0.9$,
which corresponds to the absence of current instability, and
neglect terms proportional to $\partial g_L/\partial qu$ in the
expression for $H_1(f,\xi)$. The I-V characteristics exhibits a
symmetric shape with respect to $\xi= -f/2$ as a consequence of a
$\Gamma$ which is independent of the applied voltage and of the
property of the one-dimensional density of states. From this
figure it is evident that $\gamma_{min}<0.5$ even when $f=100$. We
note, that due to the reasons discussed above, the Coulomb
interaction increases the noise for the symmetric case when
$\Gamma_L=\Gamma_R$. In addition, the reason of shot noise
enhancement is the Coulomb interaction. Accordingly,  the enhanced
result obtained in Ref. [\onlinecite{wei99}] in the absence of
Coulomb interaction is not physically plausible. From Fig. 5 it is
clear, that the Coulomb interaction in the one dimensional case
influences the noise when the resonant level is close either to
the $F_L$ ($\xi\sim -f$) or to $-g_L$ ($\xi\sim 0$).
\par
The effects due to the asymmetries of the structure are reported
in Fig. 6 which shows the dependencies of the current and Fano
factor versus $\xi$ when $\Gamma_L=0.25\Gamma$ and
$\Gamma_L=0.75\Gamma$ with the same $(C_L+C_R)$ value as in Fig.
5. We note, that in the absence of  Coulomb interaction, the Fano
factor is the same for $\Gamma_L=0.25\Gamma$ and
$\Gamma_L=0.75\Gamma$. From Fig. 6 one can see that the Coulomb
interaction decreases the value of $\gamma_{min}$. In particular,
when $\Gamma_L<\Gamma_R$, $\gamma_{min}$ is located at $\xi\sim
-f$, while, when $\Gamma_L>\Gamma_R$, at $\xi\sim 0$. We note,
that for $\xi\sim -f$ the cross correlator $<\delta I_2\delta
Q_{QW2}>$ is negative when $\Gamma_L<\Gamma_R$ and positive when
$\Gamma_L>\Gamma_R$. In both cases $\lambda$ is positive. On the
contrary, for $\xi\sim 0$ $\lambda$ is negative because $\partial
<I>/ \partial u <0$.
\par
We conclude, that also for a one dimensional RTD the
Fano factor can drop below the value 0.5
in the presence or in the absence of Coulomb interaction.
\subsection{High temperatures}
We now discuss the case of  high temperatures when $k_BT\gg\Gamma$,
which in the present case refers to $T \ge 77 \ K$.
Again, we consider applied voltages  high enough to neglect
the contribution to the current and noise of the electron flux moving
from the collector to the emitter.
Since the energy scale for the change of
$D(\varepsilon)$is significantly less than that of
$f_L(\varepsilon)$, all the integrals in the expressions for
the current and the noise spectral density can be calculated
analytically.
To perform the calculations, first of all we note that
\begin{equation}
\int_0^\infty d\varepsilon_\perp f_L(\varepsilon)=k_BT\ln\left[
1+\exp\left(\frac{F_L-\varepsilon_z}{k_BT}\right)\right]=
k_BT\Phi_L(\varepsilon_z)
\end{equation}
$$
\int_0^\infty d\varepsilon_\perp f_L(\varepsilon)^2=
$$
\begin{equation}=k_BT\left[ \Phi_L(\varepsilon_z)-\left(
1+\exp\left(\frac{F_L-\varepsilon_z}{k_BT}\right)\right)^{-1}\right]=
k_BT\Phi_{2L}(\varepsilon_z)
\end{equation}
and
\begin{equation}
\int_{-g_L}^\infty d\varepsilon_z
D(\varepsilon_z)\Phi_{L,2L}(\varepsilon_z)\approx
\frac{2\Gamma_L\Gamma_R}{\Gamma}\Phi_{L,2L}(\beta)D_1(\xi)
\end{equation}
\begin{equation}
\int_{-g_L}^\infty d\varepsilon_z
D(\varepsilon_z)^2\Phi_{L,2L}(\varepsilon_z)\approx
\frac{4\Gamma_L^2\Gamma_R^2}{\Gamma^3}\Phi_{L,2L}(\beta)D_2(\xi)
\end {equation}
here $\beta$ is the maximum value between  $(\varepsilon_r-qu)$ and $-g_L$,
$$D_1(\xi)=\frac{\pi}{2}-\arctan(\xi), \quad D_2(\xi)=D_1-\frac{\xi}{\xi^2+1}$$
By using Eqs. from (49) to (52), we can write the following
expressions, respectively for the current, the noise spectral density
and the Fano factor
\begin{equation}
<I>=\frac{qmk_BT\Gamma_L\Gamma_R}{\pi^2\hbar^3\Gamma}\Phi_L(\beta)D_1(\xi)
\end{equation}
$$
S_I(0)=\frac{2q^2mk_BT\Gamma_L\Gamma_R}{\pi^2\hbar^3\Gamma}
\left\{D_1(\xi)\Phi(\beta)-\right.
$$
$$
-\frac{2\Gamma_L\Gamma_R D_2(\xi)}{\Gamma^2}\left[
\Phi_{2L}(\beta)+\lambda
\left(\Phi_L(\beta)-\frac{2\Gamma_L}{\Gamma}\Phi_{2L}(\beta)\right)-\right.
$$
\begin{equation}
\left.\left.-\lambda^2\frac{\Gamma_L}{\Gamma}\left(\Phi_L(\xi)-\frac{\Gamma_L}{\Gamma}
\Phi_{2L}(\beta)\right)\right]\right\}
\end{equation}
$$
\gamma=1-\frac{2\Gamma_L\Gamma_R D_2(\xi)}{\Gamma^2
D_1(\xi)}\left\{\Phi(\beta)+\lambda\left(1-\frac{2\Gamma_L}{\Gamma}
\Phi(\beta)\right)-\right.
$$
\begin{equation}
\left.-\lambda^2\frac{\Gamma_L}{\Gamma}\left(1-\frac{\Gamma_L}{\Gamma}\Phi(\beta)\right)\right\}
\end{equation}
here $\Phi(\beta)=\Phi_{2L}(\beta)/\Phi_L(\beta)$.
Analogously, we derive the  expressions for $C_{QW}$ and $\partial<I>/\partial u$.
Since in this case $\partial<I>/\partial u=\Gamma_R C_{QW}/\hbar$,
we obtain
\begin{equation}
\lambda=\frac{\Gamma C_{QW}}{\Gamma_L(C_L+C_R+C_{QW})}
\end{equation}
and $\lambda<\Gamma/\Gamma_L$ in the stable region where
$(C_L+C_R+C_{QW})>0$.
\par
We now estimate the possible minimum value for $\gamma$.
From Eq. (55) it is clear that to obtain $\gamma_{min}$ it is necessary
that
\begin{equation}
\lambda=\frac{\Gamma}{2\Gamma_L}\left(1-\frac{2\Gamma_L}{\Gamma}\Phi(\beta)
\right)\left(1-\frac{\Gamma_L}{\Gamma}\Phi(\beta)\right)^{-1}
\end{equation}
We note, that this value of $\lambda$ lies in the range of possible
values characterizing the stable region.
By substituting Eq. (57) into Eq. (55) we find
\begin{equation}
\gamma_{min}=1-\frac{\Gamma_R}{2\Gamma}\frac{D_2(\xi)}{D_1(\xi)}
\left(1-\frac{\Gamma_L}{\Gamma}\Phi(\beta)\right)^{-1}
\end{equation}
Now we analyze two limiting cases.
\par
The first is the case when the occupation
factor of the state with energy $\beta$ is much smaller than unity. Then
$\Phi(\beta)\approx 0$ and:
\begin{equation}
\gamma_{min}=1-\frac{\Gamma_R}{2\Gamma}\frac{D_2(\xi)}{D_1(\xi)}
\end{equation}
Since the maximum value of $D_2(\xi)/D_1(\xi)$ equals 1.217
($\xi=\xi_0=-0.802$), from Eq. (59) we find that
%it is clear that the possibility to be
$\gamma_{min}<0.5$ when $\Gamma_R>0.821\Gamma$, and that the
expression (57) holds near $\xi=\xi_0$.
\par
The second limiting case is when the probability of occupation of an
electron state with energy $\beta$ is close to unity and
$\Phi(\beta)\approx 1$.
In this case
\begin{equation}
\gamma_{min}=1-\frac{D_2(\xi)}{2D_1(\xi)}
\end{equation}
and $\gamma_{min}<0.5$ when the expression (57) holds in any point
where $\xi<0$, since there $D_2(\xi)/D_1(\xi)>1$.
\par
In concluding this subsection, we emphasize that, for  RTD
theory predicts values of the Fano factor below 0.5 also at
high temperatures.
We note that, provided  $\Phi(\beta)=f_L(\beta)/f_L(\beta)^2$, Eq. (55)
is valid also for the one dimensional case and thus again
$\gamma_{min}<0.5$ is confirmed to be possible.
\subsection{Comparison of theory with experiments}
We compare theory with two sets of experiments performed on
DBRDs with barriers sufficiently narrow to expect that coherent tunneling
is of importance.
The first set refers to pioneer experiments of Brown\cite{brown92}
which are detailed at $77 \ K$ with indications at  $300 \ K$.
The second set refers to recent experiments at $300 \ K$ reported in
Ref. [\onlinecite{alkeev02,aleshkin03}].
In both cases the comparison is limited to the voltage region up to the
peak current since theory neglects energy levels in the quantum well
higher than the first one.
\par
Figure 7 reports the comparison between experiments performed in Ref.
[\onlinecite{brown92}] and present calculations at $77 \ K$.
Numerical results  make use of the same parameters in Fig. 4
for the symmetric structure.
For the used values, the dependence $g_L(u_L)$ is found to be almost
linear and well approximated by:
$g_L(u_L)\approx 0.44(qu_L-1.5k_B T)-0.07 k_B T$ for $77 \ K$.
% and $g_L(u_L)\approx 0.28(qu_L-110k_B T)-14 k_B T$ for $4.2 \ K$.
\par
From Fig. 7 (a) we see that present calculations well
reproduce the I-V characteristic including the current peak.
From Fig. 7 (b) we see that the calculated Fano factor
reproduces both the suppression and enhanced behaviors.
However, its  minimum value is of $0.65$ against
the experimental value of $0.35$.
By choosing appropriate values of $\Gamma_{L,R}$ the theoretical model
can be forced to fit the experimental Fano factor at the expense
of overestimating the I-V characteristics for about one order of magnitude.
%We conclude that the present theoretical model is not able
%to reproduce simultaneously the experimental I-V and noise experiments.
This result corrects previous findings of the same
authors \cite{aleshkin03}, where the contribution of charge
fluctuations to the total noise was underestimated with respect to the present approach.
In this context, we note that Ref. [\onlinecite{jahan95}]
presented a theoretical calculations of the  same
experiments \cite{brown92} at $77 \ K$.
The results of these calculations were found to be in
excellent agreement with experimental data except for the region of instability.
On the borders of this region the  noise
tends to infinity and, as a consequence, there were no measurements of
noise in this region.
However, contrary to such an experimental evidence, the theoretical
calculations \cite{jahan95} provided finite values of the noise and the
the absence of the instability region,
which makes the theoretical approach at least suspect.
\par
To provide a physical insight of the physical reason for shot-noise
suppression, Fig. 8 reports the results of the calculations associated
with the presence (continuous curves) and the absence (dotted lines)
of Coulomb interaction.
From Fig. 8 (a) we see that at $77 \ K$ the  Coulomb part contributes to
suppress shot noise in the whole region of applied voltage, and that
the Pauli  part becomes  noticeable near the current peak.
Furthermore, while Pauli contribution leads systematically to suppression,
the Coulomb contribution becomes responsible of enhanced shot noise at
voltages near to $0.57 \ V$ where the instability region is approached.
From Fig. 8 (b) we see that at $4.2 \ K$ the predominance of Pauli over
Coulomb interaction in suppressing shot noise is confirmed.
Again, Coulomb effects are found to be responsible of enhanced shot noise
around above the current peak, in agreement with experiments \cite{brown92}.
\par
To complete the analysis of this structure at $77 \ K$, we have
investigated the role played by the electron concentration and the
width of the quantum well. Accordingly, Fig. 9 reports the current
voltage characteristic and the Fano factor for the same structure
of Fig. 7 but with two different values of the electron
concentration $n$ in the injector. For the first structure we use
$g_L(u_L)\approx 0.37(qu_L-4k_B T)-1.79 k_B T$ ($n=2 \times
10^{15} \ cm^{-2}$) and for the second $g_L(u_L)\approx
0.22(qu_L-4k_B T)-0.09 k_B T$ ($n=2 \times 10^{17} \ cm^{-2}$).
From the figure it is clear that the increase of  $n$ leads to:
(i) the increase of the value of the current peak, (ii) the
widening of the region of shot noise suppression and, (iii) the
decrease of the voltage value corresponding to the  current peak
and the minimum of the Fano factor. We note that for values of $n$
which differ over two orders of magnitude the peak current changes
only for a factor of about $2.5$. It means that, due to
accumulation, the electron concentration on the border of the left
barrier differs only for a factor of $2.5$ for voltages
corresponding to the current peak. Figure 10 reports the current
voltage characteristic and the Fano factor for the same structure
of Fig. 7 but with two different values of the quantum well width
$d_{QW}$. We have taken into account that when $d_{QW}$ decreases
both $\Gamma$ and $\varepsilon_r$ increases by using for the first
structure $\varepsilon_r=145 \ meV$, $\Gamma_L=0.5 \Gamma=1 \ meV
$ and for the second structure $\varepsilon_r=45 \ meV$,
$\Gamma_L=0.5\Gamma=0.05 \ meV$. This  leads to the growth of the
current peak and its shift to higher voltages. From the same
figure we see that simultaneously with the growth of the current
peak also the width of the suppression region rises and shifts to
higher voltages. Furthermore, the minimum value of the Fano factor
is found to become more pronounced at increasing carrier
concentration and decreasing the width of the quantum well.
\par
Figure 11 reports the current voltage characteristic and the Fano
factor for the same structure of Fig. 7 at $300 \ K$. Here the solution of the Schr\"odinger equation provides:
$g_L(u_{L})=0.53(qu_{L}-1.5k_{B}T)+0.7k_{B}T$.
Calculations show, that even by increasing the temperature, the main features
of transport and noise already shown at $77 \ K$  are preserved.
However, the current peak and the minimum of the Fano factor become less
pronounced at increasing the temperature.
These trends are in agreement with experimental results\cite{brown92}
which claim a reduction of the peak-current value of about $1 \ mA$ and an
increase of the minimum of the Fano factor
when going from $77$ to $300 \ K$.
However, even at $300 \ K$ present calculations give a Fano Factor
for about a factor of two greater than that found by experiments.
\par
A recent set of experiments\cite{alkeev02,aleshkin03} carried out
at 300 K on a DBRD with barriers thinner than those of Ref.
[\onlinecite{brown92}], thus more adequate to check coherent
tunnelling at high temperature, is reported in Fig. 12 together
with theoretical calculations. The structure consisted of two $2 \
nm$ AlAs layers separated by $6 \ nm$ InGaAs quantum well
\cite{alkeev00,alkeev02}. Measurements were carried out at $300 \
K$ using a Noise Figure Meter (XK5-49), that allows to measure
simultaneously noise figure and power gain of two-port networks in
the $50 \ \Omega$ feed circuit. Simultaneously with the noise the
I-V curve was measured. The diode was mounted in the break of a
microstrip line, with one electrode been grounded, and another one
bonded to the ends of a microstrip line. Noise measurements at
frequencies $60$ and $200 \  MHz$ showed the same results within
an experimental uncertainty at worst of $20 \%$, thus indicating
that $1/f$ noise contribution  is negligible. Numerical results
makes use of the following values for the parameters entering the
model:  $\varepsilon_r=87 \ meV$, $\Gamma_R=\Gamma_L=1.08 \ meV$,
$n=5\times 10^{16} \ cm^{-3}$, $d_L=d_R=2 \ nm$, $d_{QW}=6 \ nm$
$L=50 \ nm$, $g_L(u_{L})=0.445(qu_{L}-1.5k_{B}T)+0.451k_{B}T$,
$m=0.045m_0$.
Also here the only two fitting parameters are
$\varepsilon_r$ and $\Gamma=0.5\Gamma_{L,R}$, all other parameters
being provided by the experimental conditions, Figure 12(a) reports
the I-V characteristic which shows a region of positive
differential conductance (pdc) up to about $0.7 \ V$ followed by
an instability region. In the same pdc region, the Fano factor is
found to exhibit a suppression with a minimum value of about $0.4$
at around $0.65 \ V$ [see Fig. 12(b)].
As for the case at $77 \ K$, from Fig. 12 (a) we see that present
calculations well reproduce the I-V characteristic including the current peak.
From Fig. 12 (b) we see that the calculated Fano factor
reproduces both the suppression and enhanced behaviors.
However, the minimum value of the Fano factor is found to be of $0.75$
against the experimental value of $0.40$.
By choosing larger values of $\Gamma_{L,R}$ the theoretical model
can be forced to fit the Fano factor at the expenses
of overestimating the I-V characteristic for about one order of
magnitude.
This result corrects previous findings of the same authors \cite{aleshkin03},
where the contribution of charge fluctuations to the total noise
was underestimated.
Thus, the comparison between theory and experiments at temperatures above
about $77 \ K$ provides qualitative agreement but is not able to
explain the drop of the Fano factor below $0.5$ found in experiments.
\section{Conclusions}
We have implemented a quantum mechanical approach to investigate DBRDs
transport and noise characteristics within the coherent tunneling
regime that includes both Pauli principle and long range Coulomb interaction.
The expression for the current voltage and noise characteristics generalize
previous findings\cite{averin93,blanter99,wei99}.
In agreement with expectations, at increasing voltages theory predicts
a current characteristic which exhibits a peak followed by an instability region and that before the current peak shot noise is suppressed because of
Pauli principle and/or Coulomb interaction.
In addition, theory confirms shot noise enhancement well above the full
Poissonian value at the current peak as a consequence of the positive
feedback between tunneling and space charge.
\par
At zero temperature, the suppression of shot noise  starts in
concomitance with the sharp increase of the current associated
with the alignment within the value of $\Gamma$ of the Fermi level
in the emitter with the resonant level in the quantum well.
Accordingly, the Fano factor is found to exhibit a minimum  near
the current peak. Remarkably, the value of this minimum can be
significantly below the value $0.5$ of the full Poissonian value.
This giant suppression of shot noise  can be taken as a signature
of coherent tunneling since sequential tunneling can predict
suppression but at most with a Fano factor not less than 0.5. At
$4.2 \ K$, for a realistic DBRD structure we find a minumum value
of the  Fano factor of 0.44 that is in agreement with experiments
\cite{brown92}. At temperatures above about $77 \ K$, we have
found that coherent tunneling still predict that shot noise can be
suppressed well below the value of $0.5$. This giant suppression
has been evidenced by experiments performed at $77$ and $300 \ K$.
However, the present theory is not able to explain this noise
feature with the same set of parameter values able to explain the
I-V characteristics. Probably the model used is still too simple,
and further efforts are needed to provide a better interpretation
of experiments.
\par
In any case, to our opinion, the main result obtained here is the
fact that in RTDs the drop of the Fano factor below 0.5 can be
taken as a signature of coherent against sequential tunneling.
Therefore, below we briefly discuss  the physical reason why shot
noise suppression is expected to be more effective for coherent
than for sequential tunneling. Starting from the fact that the two
mechanisms responsible for shot noise suppression are Pauli
principle and Coulomb interaction, we note that  both act
simultaneously for coherent and sequential tunneling. Let us
consider the first mechanism, which is the most relevant at zero
temperature, in the case when the Fermi energy is sufficiently
small so that all the electrons exhibit the same transparency.
Then, coherent transport admits a transparency near equal to
unity, which implies $\gamma=(1-D) \simeq 0$ according to Lesovik
findings\cite{lesovik89}. For sequential transport the total
transparency is always less than unity because of the finite value
of the differential rates controlling the relaxation of carrier
number fluctuations inside the two terminal device through the
contacts.
%due to scattering an electron can change
%the direction of  motion.
As a consequence,  under coherent transport for the possible case
of full transparency (i.e. $D=1$) there is no noise.
By contrast, under sequential transport the presence of
scattering inside the quantum well always introduces noise.
This example illustrates why Pauli principle is more efficient in
suppressing shot noise under coherent than sequential transport.
By considering Coulomb interaction, which is more relevant
at high temperatures, we recall that in the absence of collisions
it provides giant shot noise suppression\cite{gomila02} as in the vacuum tubes\cite{ziel54}
because electron reflection in this case is due only to
Coulomb interaction.
It is clear that the presence of scattering provides additional
random mechanisms for electrons returning to the emitter and,
therefore, provides an additional source of noise.
Even this example shows that Coulomb interaction is more efficient
in suppressing shot noise under coherent than sequential transport.
We finally want to stress that the main reason of the
difference between these approaches stems from the fact that the
sequential tunneling  is based on a master
equation\cite{davies95,iannaccone97} for treating fluctuations of
carrier numbers inside the quantum well, while coherent tunneling
uses the quantum partition noise as only source of interaction and
fluctuations. The master equation describes implicitly
a sequential mechanism for a carrier entering/exiting from the
well and, as a consequence, its intrinsic limit coincides with
that of two independent resistors (or vacuum diodes)  connected
in series and each of them exhibiting full shot noise.
This system yields a maximum
suppression of shot noise down to the value of 0.5.
By contrast, partition noise,  inherent to a quantum coherent
formalism, can be fully suppressed down to zero in the presence of
a fully transparent barrier and weak Coulomb interaction like in
vacuum diodes.

\vskip1pc\noindent
ACKNOWLEDGMENTS
\vskip1pc\noindent
We thank Prof. V. Volkov of Moscow Institute of Radioengineering and
Electronics for having suggested the problem of this research and Prof. M.
B\"uttiker of Geneva University for having addressed the importance of
charge fluctuations in the calculation of the current fluctuations.
This research has ben performed within the project "Noise models and measurements in nanostructures" supported by the Italian Ministery of Education, University and Research (MIUR) through the cofin03.
Partial support from the Italian Ministry of Foreign Affairs
through the Volta Landau Center (the fellowship of V.Ya.A.),
and the SPOT-NOSED project IST-2001-38899 of the EC is
gratefully acknowledged.

%

%
%\newpage
%
\section{Appendix}

Here we detail the calculations to evaluate the current operator,
the differential capacitances of the DBRDs and the solution of theSchr\"odinger equation for the voltage dependence of the first
electron level in the emitter.
\subsection{Current operators}

By using Eqs. (3), (4), (5) and (7), the explicit expression for $A^\alpha_{\beta,\gamma}(E_\beta,E_\gamma)$ is the following
$$
A^\alpha_{\beta,\gamma}(E_\beta,E_\gamma)=-2\pi i\left(
\frac{T_\alpha
T_\beta^*B_{\alpha\gamma}(E_\gamma)}{E_\beta-E_r-i\Gamma/2}-\right.
$$
$$
\left.-\frac{T_\alpha^*T_\gamma
B^*_{\alpha\beta}(E_\beta)}{E_\gamma-E_r+i\Gamma/2}\right)
\eqno(A1)
$$
where
$$
B_{\alpha\beta}(E_\beta)=\left(\delta_{\alpha\beta}-i\pi
\frac{T_\alpha^*T_\beta\rho_\alpha}{E_\beta-E_r+i\Gamma/2}\right)\eqno(A2)
$$
From the above, we note the following properties:
$$
A^L_{LL}(E,E)\rho_L=A^R_{RR}(E,E)\rho_R=-A^L_{RR}(E,E)\rho_R=
$$
$$
=-A^R_{LL}(E,E)\rho_L=D(E)\eqno(A3)
$$
and
$$
A^L_{LR}(E,E)A^L_{RL}(E,E)\rho_L\rho_R=
$$
$$A^R_{LR}(E,E)A^R_{RL}(E,E)\rho_L\rho_R
=D(E)(1-D(E))\eqno(A4)
$$
Eqs. (A3) and (A4) are useful for current and noise calculations.
\subsection{Capacitances}
By recalling that $Q_{L}=-\kappa
%\kappa_0
\varphi _{L}^{\prime }/4\pi $ from Eq. (12) we have
$$
C_{e}(u_{L})=q\left\{N_{c}F_{1/2}\left( \frac{F_{L}+
qu_{L}}{k_{B}T}\right) -n\right\} \sqrt{\frac{\kappa }{8 \pi
k_{B}T}}\times
$$
$$
\left[ N_{c}F_{3/2}\left( \frac{%
F_{L}+qu_{L}}{k_{B}T}\right) -N_{c}F_{3/2}\left(
\frac{F_{L}}{k_{B}T}\right) -\frac{q}{k_{B}T}nu_{L}\right]^{-1/2}
\eqno(B1)
$$
If the electron accumulation is relevant (i.e. $qu_{L}>k_BT$) then we can
write
$$
C_{e}(u_{L})\approx q\frac{N_{c}F_{1/2}\left( \frac{F_{L}+qu_{L}}
{k_{B}T}\right) }{\sqrt{\frac{8\pi k_{B}T}{\kappa \kappa_0
}N_{c}F_{3/2} \left( \frac{F_{L}+qu_{L}}{k_{B}T}\right) }}
\eqno(B2)
$$
By substituting for $Q_{R}$ the value given in Eq. (15), for
$C_{c}$ we find
$$
C_{c}=q\left\{n-N_{c}F_{1/2}\left(
\frac{F_{L}+qu_{L}-qV}{k_{B}T}\right)\right\} \sqrt{\frac{\kappa
}{8\pi k_{B}T}} $$
$$ \left[
N_{c}F_{3/2}\left( \frac{F_{L}+qu_{R}-qV}{k_{B}T}\right)
-N_{c}F_{3/2}\left( \frac{F_{L}-qV}{k_{B}T}\right)\right.
$$
$$
\left.-\frac{qn}{k_{B}T}\left( u_{R}-V\right) \right]^{-1/2}
\eqno(B3)
$$
If $q(V-u_{R})/(k_BT)>1$, then we have the usual expression for the
capacitance of the depletion region
$$
C_{c}\approx \sqrt{\frac{q\kappa  n}{8\pi (V-u_{R})}} \eqno(B4)
$$
\subsection{Energy of the first electron level in the emitter}
The one dimensional Schr\"odinger equation for an electron moving
in the potential $-q\varphi (x) $ can be written in the following form
$$
\frac{dy}{dx}+y^{2}+\frac{2m}{\hbar ^{2}}\left( q\varphi
+\varepsilon \right) =0\eqno(C1)
$$
here $y=\Psi ^{\prime }/\Psi $, $\Psi $ is the electron wave
function and $\varepsilon $ the electron energy. Since $d\varphi
/dx$ in the emitter is function of $\varphi $ [see Eqs.  (11) and
(12)], instead of $x$ it is convenient  to use $\varphi $ asvariable quantity so that
$$
\frac{dy}{d\varphi }\frac{d\varphi }{dx}+y^{2}+\frac{2m}{\hbar
^{2}}\left( q\varphi +\varepsilon \right) =0\eqno(C2)
$$
where
$$
\frac{d\varphi }{dx}=\sqrt{\frac{8\pi k_{B}T}{\kappa }}\left[
N_{c}F_{3/2}\left( \frac{F_{L}+q\varphi }{k_{B}T}\right)
-N_{c}F_{3/2}\left( \frac{F_{L}}{k_{B}T}\right)\right.
$$
$$
\left.-\frac{q}{k_{B}T}n\varphi \right]^{1/2}\eqno(C3)
$$
The potential $\varphi $ in the emitter takes values in the range
from $0$ far from the barriers $ (x \rightarrow -\infty )$
to $u_{L}$ on the border of the left barrier.
We note that far from the barriers the potential falls
exponentially $\varphi \sim \exp (x/\lambda _{D})$ when $q\varphi
\ll k_{B}T$, where $\lambda _{D}$ is the Debye screening length. For
localized state the electron wave function $\Psi \sim \exp
(k_{1}x)$ for $x \rightarrow -\infty $, where $k_{1}=\sqrt{
-2m\varepsilon }/\hbar $ and, thus, we have as boundary condition
$y(\varphi =0)=k_{1} $.
Since we suppose that the field in the barrieris absent, the electron
wave function in the barrier is $\Psi \sim \exp (-k_{2}x)$
where $k_{2}=\sqrt{%
2m\left( \Delta E_{c}-qu_{L}-\varepsilon \right) }/\hbar $ and
$\Delta E_{c}$ is the conduction band offset on the barrier
border. By using the conditions of continuity of the wave function
and of its derivative, we obtain the second boundary condition
$y(\varphi =u_{L})=-k_{2}$. Equations (C2) and (C3) together with
the boundary conditions allow us to find the energy of the first
electron level in the potential well of the emitter by numerical
calculations.

\newpage
\begin{figure}
%\centerline{\includegraphics[width=1.8\linewidth]{f1b.eps}}
\includegraphics[72,180][701,601]{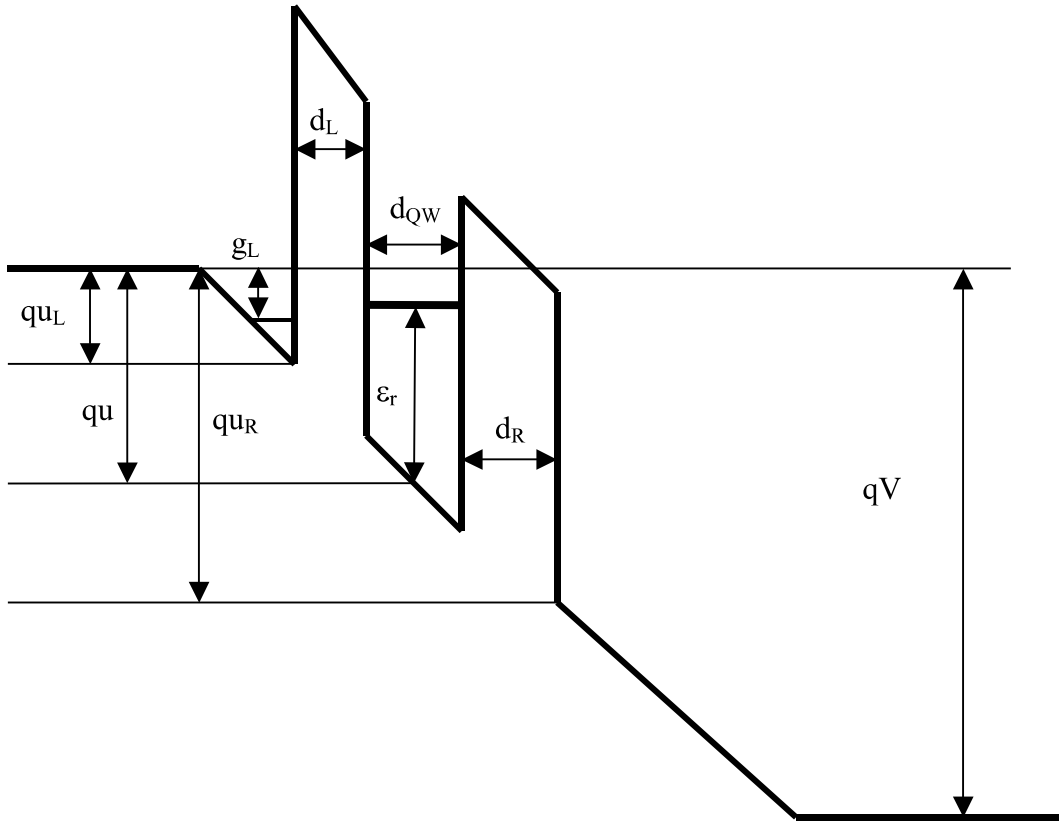}
\caption{Sketch of the band profile of the double barrier
structure considered here The bottom of the conduction band in the
emitter in the well and in the collector coincides at $V=0$}
\end{figure}

\begin{figure}
\centerline{\includegraphics[width=1.2\linewidth]{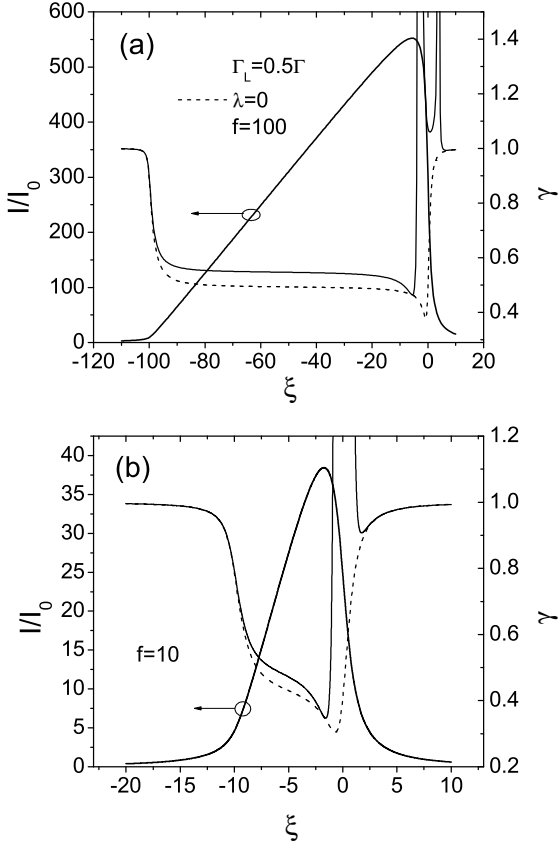}}
\caption{Dependence of the Fano factor and of the current on the
electrical potential in dimensionless units
$\xi=2(qu-\varepsilon_r-g_L)/\Gamma$ with $f=100$ (a) and $f=10$
(b) for the symmetrical structure $\Gamma_L=\Gamma_R$. Here
$f=2(F_L+g_L)/\Gamma$ and $I_0=qm\Gamma_L \Gamma_R/(2\pi^2
\hbar^3)$. Continuous (dashed) curves correspond to the presence
(absence) of Coulomb interaction.}
\end{figure}

\begin{figure}
\centerline{\includegraphics[width=1.2\linewidth]{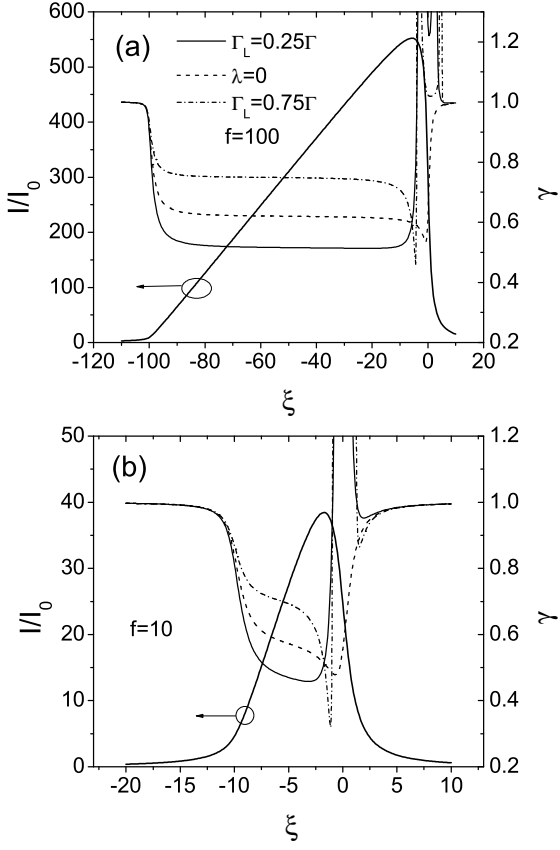}}
\caption{ Dependence of the Fano factor and of the current on the
electrical potential for asymmetric structures with $f=100$ (a)
and $f=10$ (b). The dimensionless units are the same of Fig. 2.}
\end{figure}
\begin{figure}
\centerline{\includegraphics[width=1.2\linewidth]{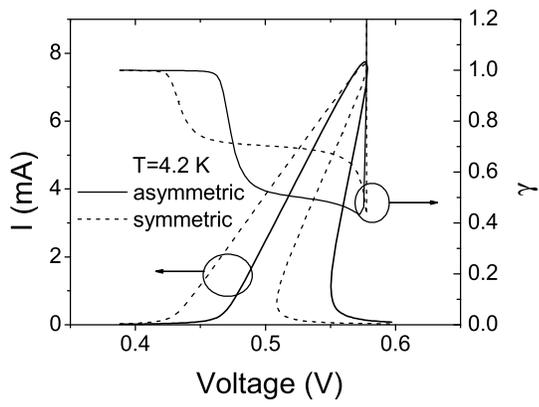}}
\caption{Dependence of the current and Fano factor on the applied
voltage for the structure of Ref. [\cite{brown92}] at $T=4.2 \ K$.
Values of $\Gamma_{L,R}$ and $\varepsilon_r$ are chosen from the
fitting of the current voltage characteristic at $77 \ K$.}
\end{figure}

\begin{figure}
\centerline{\includegraphics[width=1.2\linewidth]{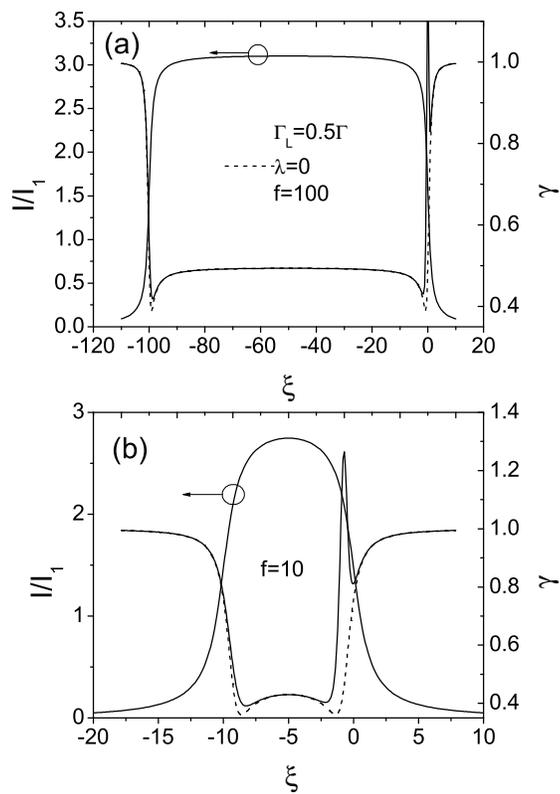}}
\caption{Dependence of the Fano factor and current on the
electrical potential for a one-dimensional symmetric structure
$\Gamma_L=\Gamma_R$ with $f=100$ (a), and $f=10$ (b). Continuos
(dashed) curves correspond to the presence (absence) of Coulomb
interaction. Here $I_1=2q\Gamma_L \Gamma_R/(\pi\hbar\Gamma)$.}
\end{figure}

\begin{figure}
\centerline {\includegraphics[width=1.2\linewidth]{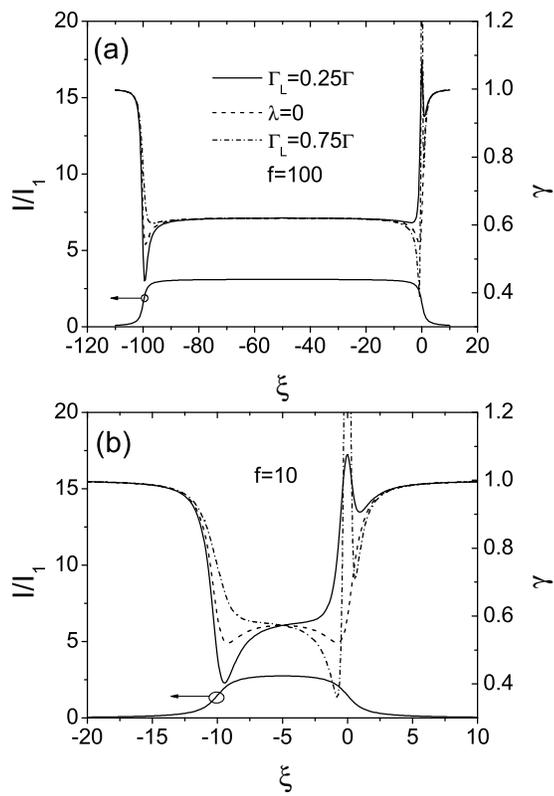}}
\caption{Dependence of the Fano factor and current on the
electrical potential  for asymmetrical one dimensional structures
with $f=100$ (a), and $f=10$ (b)}
\end{figure}

\begin{figure}
\centerline{\includegraphics[width=1.2\linewidth]{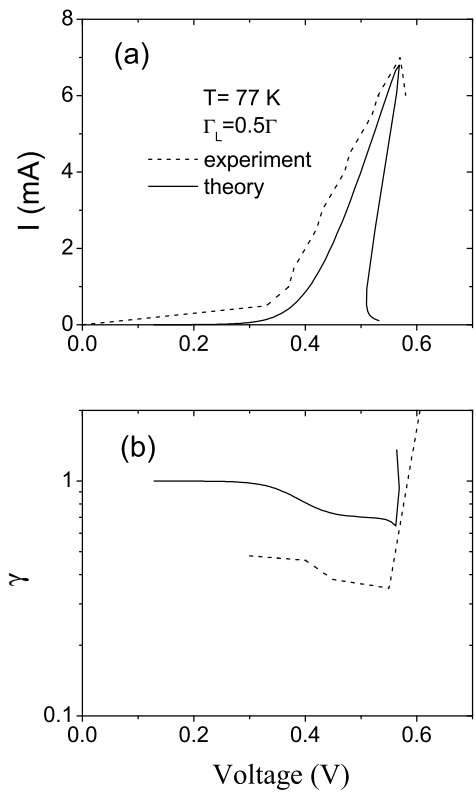}}
\caption{Calculated (solid) and measured (dashed) dependencies of
current and Fano factor on the applied voltage for the structure
of \cite{brown92} at $77 \ K$. The parameters are the same used
for Fig. 4 in the case of the symmetric structure. }
\end{figure}

\begin{figure}
\centerline{\includegraphics[width=1.2\linewidth]{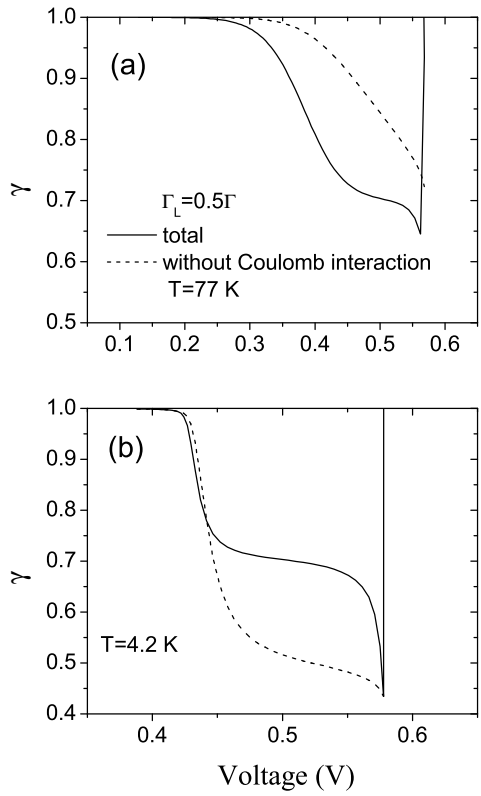}}
\caption{Calculated dependencies of the Fano factor with
(continuous curves) and without (dashed curves) Coulomb
interaction at $ 4.2 $ and $77 \ K$. Other parameters are as in
Fig. 7.}
\end{figure}

\begin{figure}
\centerline{\includegraphics[width=1.2\linewidth]{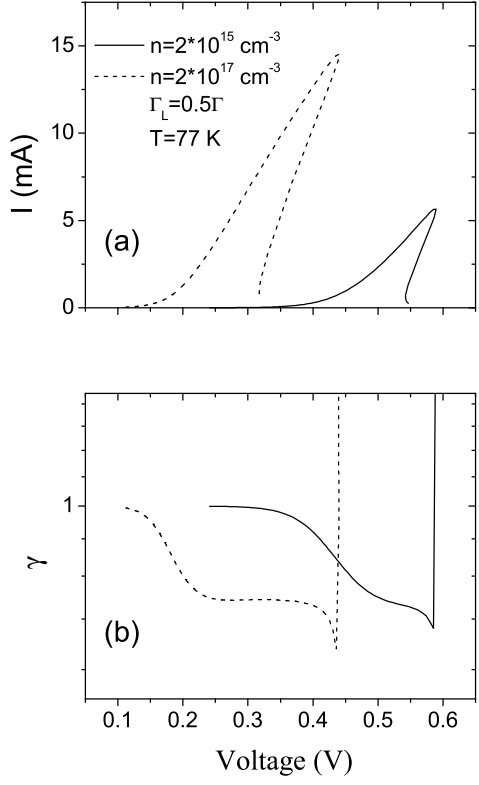}}
\caption{ Dependencies of current and Fano factor on applied
voltage for structures with $n=2\times 10^{15}$~cm$^{-3}$ (solid)
and $n=2\times 10^{17}$~cm$^{-3}$ (dashed). Other parameters are
as in Fig. 7.}
\end{figure}

\begin{figure}
\centerline{\includegraphics[width=1.2\linewidth]{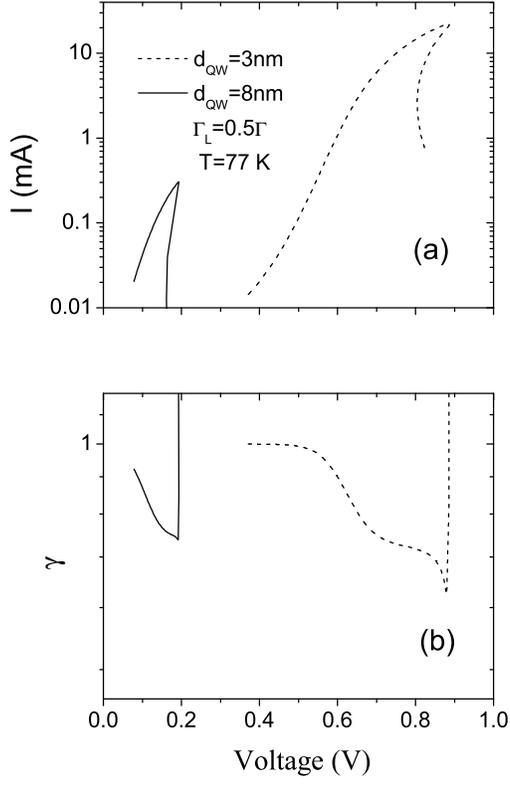}}
\caption{ Dependencies of current and Fano factor on applied
voltage for structures with $d_{QW}=3$~nm (dashed curve) and
$d_{QW}=8$~nm (solid curve). For the first structure we used
$\varepsilon_r=145$~meV, $\Gamma_L=0.5 \Gamma=1$~meV. For the
second structure we used $\varepsilon_r=45$~meV,
$\Gamma_L=0.5\Gamma=0.05$~meV. Other parameters are as in Fig. 7.}
\end{figure}

\begin{figure}
\centerline{\includegraphics[width=1.2\linewidth]{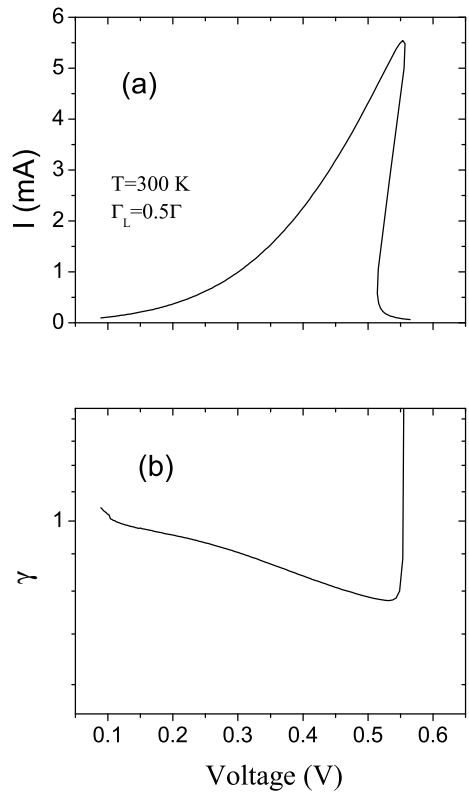}}
\caption{ Dependencies of current and Fano factor on applied
voltage at $T=300 \ K$. Other parameters are as in  Fig. 7.}
\end{figure}

\begin{figure}
\centerline{\includegraphics[width=1.2\linewidth]{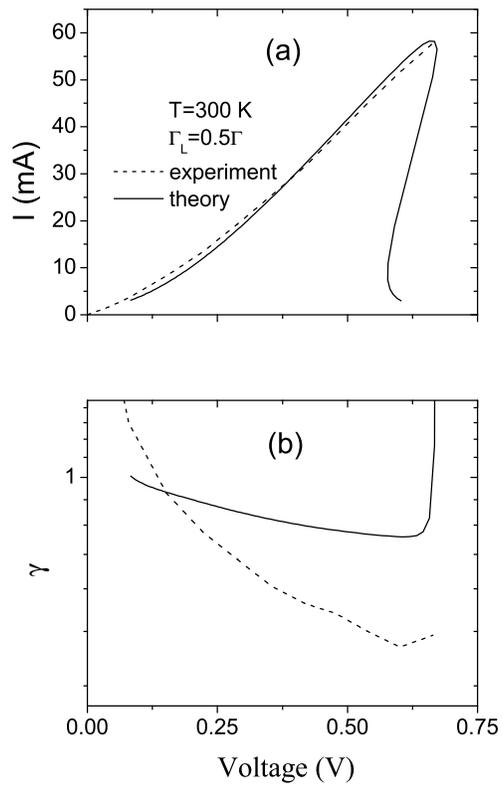}}
\caption{ Experimental and calculated dependencies of current (a)
and Fano factor (b) on applied voltage for the structure in Ref.
[\onlinecite{alkeev02,aleshkin03}] at $T=300 \ K$.}
\end{figure}

\end{document}